\documentclass[namedreferences]{solarphysics}

\usepackage[hyperref,optionalrh]{spr-sola-addons} 
\usepackage[optionalrh]{spr-sola-addons} 
\usepackage{epsfig}          
\usepackage{graphicx}        
\usepackage{epstopdf}
\usepackage[pdftex]{lscape}
\usepackage{amssymb}        
\usepackage{color}           
\usepackage{multirow}
\usepackage{booktabs}

\chardef\us=`\_
\begin{document}

\begin{article}
\begin{opening}

\title{Interferometric Observations of the Quiet Sun at 20 and 25 MHz in May 2014.\\}


\author[addressref={aff1},corref,email={melnik@rian.kharkov.ua}]{\inits{V.N.}\fnm{V.N.}~\lnm{Melnik}}
\author[addressref=aff1,email={shepelev@rian.kharkov.ua}]{\inits{V.A.}\fnm{V.A.}~\lnm{Shepelev}}
\author[addressref=aff2,email={Stefaan.Poedts@wis.kuleuven.be}]{\inits{S.}\fnm{S.}~\lnm{Poedts}}
\author[addressref=aff1]{\inits{F.}\fnm{V.V.}~\lnm{Dorovskyy}}
\author[addressref=aff3]{\inits{F.}\fnm{A.I.}~\lnm{Brazhenko}}
\author[addressref=aff4]{\inits{F.}\fnm{H.O.}~\lnm{Rucker}}


\address[id=aff1]{Institute of Radio Astronomy, Kharkov, Ukraine}
\address[id=aff2]{Catholic University of Leuven, Leuven, Belgium}
\address[id=aff3]{Institute of Geophysics, Gravimetrical Observatory, Poltava, Ukraine}
\address[id=aff3]{Commission for Astronomy, Graz, Austria}

\runningauthor{V.N. Melnik et al.}
\runningtitle{Interferometric Observations of the Quiet Sun at 20 and 25 MHz in May 2014.}

\begin{abstract}
Results of solar observations at 20 and 25 MHz by the UTR-2 (\textit{Ukrainian T-shaped Radio telescope of the second modification}) radio telescope  in the interferometric session from 27 May to 2 June 2014 are presented. In such a case the different baselines 225, 450, and 675 m between sections of East--West and North--South arms of the radio telescope UTR-2 were used. On 29 May 2014,  strong sporadic radio emission consisting of Type III, a Type II and a Type IV bursts was observed. On other days there was no solar radio activity in the decameter range.
We discuss the results of observations of such the quiet Sun. Fluxes and sizes of the Sun in East--West and North--South directions  were measured.  The average fluxes were 1050--1100 Jy and 1480--1570 Jy at 20 and 25 MHz, respectively. Angular sizes of the quiet Sun in equatorial and polar directions were $55'$ and $49'$ at 20 MHz and  $50'$ and  $42'$ at 25 MHz. Brightness temperatures of radio emission were ${T_{\rm b}} = 5.1 \times {10^5}$ K and ${T_{\rm b}} = 5.7 \times {10^5}$ K  at 20 and 25 MHz, respectively.
\end{abstract}
\end{opening}

\section{Introduction}
     \label{S-Introduction}

The large size of the radio telescope UTR-2 (\textit{Ukrainian T-shaped Radio telescope of the second modification}) allowed us to use it as an interferometer with baselines between its sections up to 1.6 km in the N--S direction and up to 700 m in the E--W direction \citep{Shepelev15,Melnik2017}. Such interferometers  provide the  opportunity to define the location of radio emission sources and their sizes at decameter wavelengths  if they have sizes from some minutes to tens of minutes. Such sizes are close to those of the quiet Sun \citep{Aubier71,Kundu77,Erickson77,Bhonsle79,Abranin86,Thejappa92,Sastry94,Subramanian04,Ramesh06,Brazhenko12,Stanislavsky13} and the sources of Type III bursts \citep{Abranin76,Chen78,Bhonsle79,Abranin80,Suzuki85,Melnik2017}, Type II bursts \citep{Chen78,Bhonsle79,Nelson85}, and Type IV bursts \citep{Gergely74,Chen78} in the decameter range. Previously the sizes of these bursts were measured only occasionally at some discrete frequencies, but now we have the opportunity to observe radio emission of different bursts in the whole frequency band from 8 to 33 MHz on a regular basis.

During the period from 27 May to 2 June 2014 there was only one powerful CME on 29 May, accompanied by groups of Type III bursts, a Type II burst and a Type IV burst, which were observed by the radio telescopes UTR-2 and URAN-2 (\textit{Ukrainian Radio interferometer of Academy of Scienses}). The frequency, temporal, and spatial properties of these bursts will be discussed in a follow-up article. On other days,  27, 28, 30, 31 May, and 2 June, the activity of the Sun was low, and only separate weak Type III bursts were observed. This opened the opportunity to study the radio emission of the quiet Sun before and after a CME.

In this article results of interferometric observations of the quiet Sun at frequencies 20 and 25 MHz are presented. Radio emission of the quiet Sun in the wide frequency band had been studied since the 1940s. Mainly it dealt with mm-, cm-, dm-, and m- wavelength bands. There were only rare observations of the quiet Sun at frequencies slightly higher and lower than 30 MHz. The article by Aubier, Leblanc, and Boishot (1971) was one of the first, in which radio emission of the quiet Sun at decameter wavelengths was analyzed. \citeauthor{Aubier71}  (\citeyear{Aubier71}) observed the Sun with the Arecibo radio telescope and obtained fluxes of $3300 \pm 500$ Jy and $4000 \pm 600$ Jy at frequencies 29.3 and 36.9 MHz, respectively. The angular sizes of the Sun in the equatorial plane were $70' \pm 10'$  and $53' \pm 5'$ ; the brightness temperatures equaled $3.8 \times {10^5}$ K and $5 \times {10^5}$ K at these frequencies.

In May\,--\,June 1976 and April 1977 \citeauthor{Kundu77}  (\citeyear{Kundu77}) observed the Sun by means of the radio telescope at Clark Lake at 26.3 MHz and found the fluxes, sizes, and brightness temperatures of the quiet Sun radio emission to be equal to $1.6 \pm 0.4 \times {10^3}$ Jy, $56.6' \pm 6'$, $4.7 \times {10^5}$ K, respectively.  Observations of the quiet Sun performed by \citeauthor{Erickson77}  (\citeyear{Erickson77}) with this radio telescope in July\,--\,August 1977 gave fluxes  $ < 1.1 \times {10^3}$ Jy, $1.5 \pm 0.3 \times {10^3}$ Jy, $1.2 \pm 0.2 \times {10^3}$ Jy, sizes $55'$, $50'$, $48'$ and brightness temperatures  $5 \times {10^5}$ K, $4.4 \times {10^5}$ K, $2.6 \times {10^5}$ K at frequencies 19, 25.8, and 30.9 MHz, respectively. The average angular sizes of the quiet Sun in East--West and North--South directions obtained by the radio heliograph at Clark Lake in September 1986 \citep{Thejappa92} equaled  $64'$ and $53'$, while the brightness temperatures varied from $0.5 \times {10^5}$ K  to $2.5 \times {10^5}$ K at 38.5 MHz.

Observing the Sun with the UTR-2 heliograph in 1976\,--\,1977 \citeauthor{Abranin86}  (\citeyear{Abranin86}) found smaller fluxes of 770--990 Jy and sizes $56'$--$58'$ at frequency 25 MHz.

At a frequency of 34.5 MHz the quiet Sun was studied many times with the radio telescope in Gauribidanur (India). Observations in August 1983 \citep{Sastry94} showed that the diameter of the quiet Sun was 3\,--\,4 R$_\odot$  and the brightness temperature changed from day to day in the range $1 \times {10^5}$ \,--\,$7 \times {10^5}$ K. \citeauthor{Subramanian04}  (\citeyear{Subramanian04}) observed the angular size in the equatorial plane from $39'$  to  $66'$ and brightness temperature from $1 \times {10^5}$ K to $4.5 \times {10^5}$ K during the observations in June\,--\,July 1986 and in May\,--\,June 1987.

During the last solar minimum (2008\,--\,2009) the radio emission of the quiet Sun was observed by the radio telescope URAN-2 (Poltava, Ukraine) \citep{Brazhenko12}. The fluxes obtained were relatively small and equaled  710  and 870 Jy  in 2008 and 700 and 860 Jy in 2009 at frequencies 20 and 25 MHz, respectively. In September 2010 the quiet Sun was observed by the UTR-2 radio heliograph \citep{Stanislavsky13}, which gave fluxes of  763 and 1094 Jy  and (E--W)$\times$(N--S) sizes of $60' \times 50'$  and $58' \times 42'$ at the same set of frequencies.

We see that for the periods of deep minimum of solar activity and for the years of its increasing phase the measured fluxes, sizes, and brightness temperatures differ significantly. One can say that parameters of the quiet Sun are non-stationary. It is of interest to analyze in what way the sizes of the Sun and its brightness temperatures vary during the 11-year cycle and how these parameters are affected by such strong phenomena as CMEs, which disturb the solar corona. For this reason regular observations of the Sun by UTR-2 in interferometric mode seem to be very promising because of its high sensitivity, large sizes, and wide working frequency range.

In this article we consider the first results obtained in solar observations by UTR-2 in the interferometric regime with different baselines in East--West and North--South directions at frequencies 20 and 25 MHz aimed to obtain reliable data concerning the time before and after CME, which occurred on 29 May 2014.

\section{Radio Telescopes}
\label{S-telescopes}

The UTR-2 radio telescope is situated near Kharkiv (Ukraine). This is a T-shape antenna array with two arms, the East--West and North--South. The East--West arm has a length of 900 m and a width of 39.3 m and consists of 600 dipoles (Figure 1). The North-South arm with length of 1860 m and width of 53 m consists of 1440 dipoles \citep{Braude78,Konovalenko16}. All dipoles are grouped in four (East--West) and eight (North--South) sections, which can be operated separately. The beam of each section is ${4^ \circ } \times {15^ \circ }$  and the beam of the whole antenna is $25' \times 25'$ at a frequency 25 MHz. The total area of the radio telescope comprises about  140,000 m$^2$.  The working frequency band of the UTR-2 is 8\,--\,33 MHz.  The spectral analyzers at the interferometric observations are operating at frequencies 20 and 25 MHz with frequency width of 250 kHz.

Observations in the interferometric regime by the UTR-2 radio telescope were supported by observations of the URAN-2 radio telescope.

The URAN-2 radio telescope is near Poltava (Ukraine) \citep{Brazhenko05}. This is a rectangular array with an effective area of 28,000 m$^2$  operated in the frequency band of 8\,--\,33 MHz. This antenna array has a size of 238 m in East--West direction and 118 m in North--South direction and a beam size of ${3.5^ \circ } \times {7^ \circ }$ at a frequency of 20 MHz. The recording of signals is performed with a digital spectrum analyzer (DSPz) \citep{Ryabov10}, which allows us to carry out observations with time--frequency resolution of 4 kHz -- 100 ms, and dynamic range of 90 dB in the frequency band of 8\,--\,32 MHz.

\begin{figure}    
   \centerline{\hspace*{0.015\textwidth}
               \includegraphics[width=0.415\textwidth,clip=]{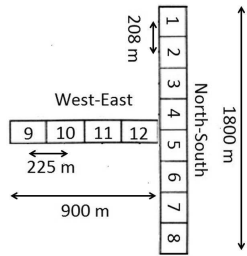}
               }

\caption{Layout of the UTR-2 radio telescopes.
        }
   \label{F1}
   \end{figure}

Because sections of UTR-2 have independent outputs there is an opportunity to shape interferometers with different baselines. We used baselines between neighboring sections (225 m), over one section (450 m), and over two sections (675 m) of the East--West and North--South arms.

The angular size of a radio emission source is defined  by the  analysis of the visibility function \citep{Thompson01}. Figure 2 shows visibility functions [$\gamma $]  of sources with different sizes for interferometers with different baselines [$L$] at 25 MHz according to the equation

\begin{equation}  \gamma  = \exp [ - {(\frac{{\pi \theta L}}{{2\lambda \sqrt {\ln 2} }})^2}]
   \end{equation}
where  $\theta $ is the half flux angular size and $\lambda $  is the wavelength at which observations are carried out. It can be seen that sources with angular sizes from some minutes to tens of minutes can be measured by the interferometers of UTR-2. Since the sizes of the quiet Sun at decameter wavelengths equal tens of minutes and sources of bursts are less than or equal to those of the quiet Sun \citep{Bhonsle79} then such interferometers can be effective for measurements of sizes in the wide frequency band from 8 to 32 MHz.

 \begin{figure}    
   \centerline{\includegraphics[width=0.7\textwidth,clip=]{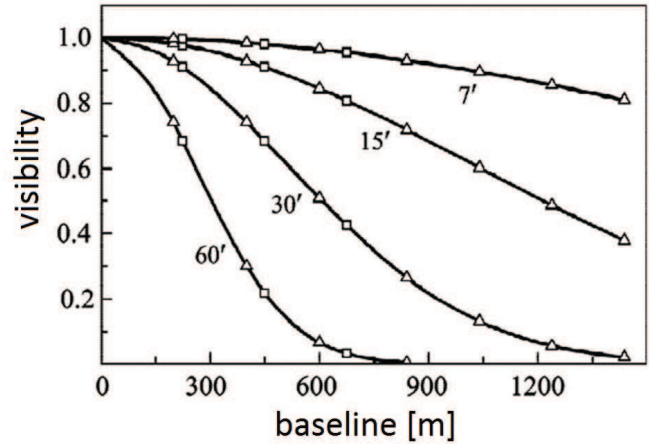}
              }
              \caption{ Dependences of visibility functions for Gaussian sources of different sizes \emph{versus} baseline of UTR-2 interferometers in E--W (squares) and N--S (triangles) directions at 25 MHz.
                      }
   \label{F2}
   \end{figure}

\section{Observations}
\label{S-obs}

Observations were carried out from 27 May to 2 June 2014 by ten minute scans at angles $ \pm $ two  hours from culmination. For calibration the source 3C144 was used. In the observational period this source had the same declination and one hour separation in right ascension from the Sun, so after every two solar scans the interferometer beam was directed to the 3C144.

As has been said above, these observations were supported also by observations with the URAN-2 radio telescope in the frequency band of 8\,--\,32 MHz. The dynamic spectra of solar radio emission on the day before and the day after the CME ejected are presented in Figure 3. It can be seen that on these days there was no  important radio activity in the decameter range. At the same time there were some active regions on the solar disk (Figure 4). The CME on 29 May 2014 was ejected in the north east direction and was accompanied by groups of Type III, a Type II, and a Type IV bursts observed by both UTR-2 and URAN-2 radio telescopes. Consideration of their properties including sizes of their sources and their locations will be presented in the next article. Here we discuss the radio emission fluxes of the quiet Sun and its sizes before and after the CME.

 \begin{figure}    
   \centerline{\includegraphics[width=0.99\textwidth,clip=]{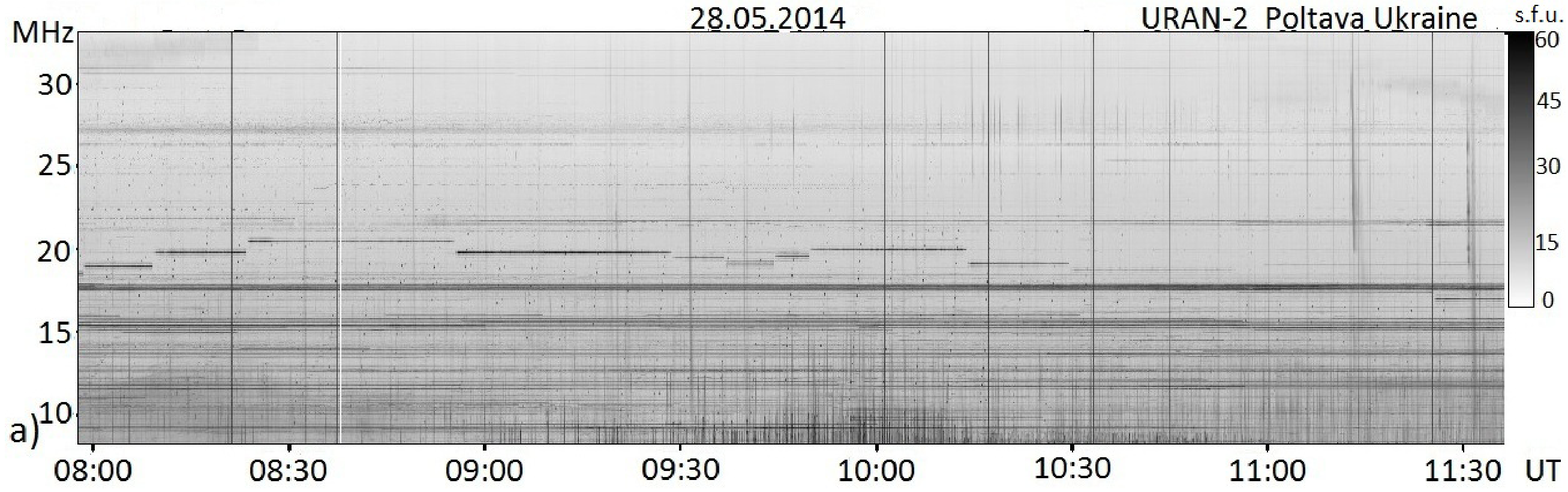}
                             }

   \centerline{ \includegraphics[width=0.99\textwidth,clip=]{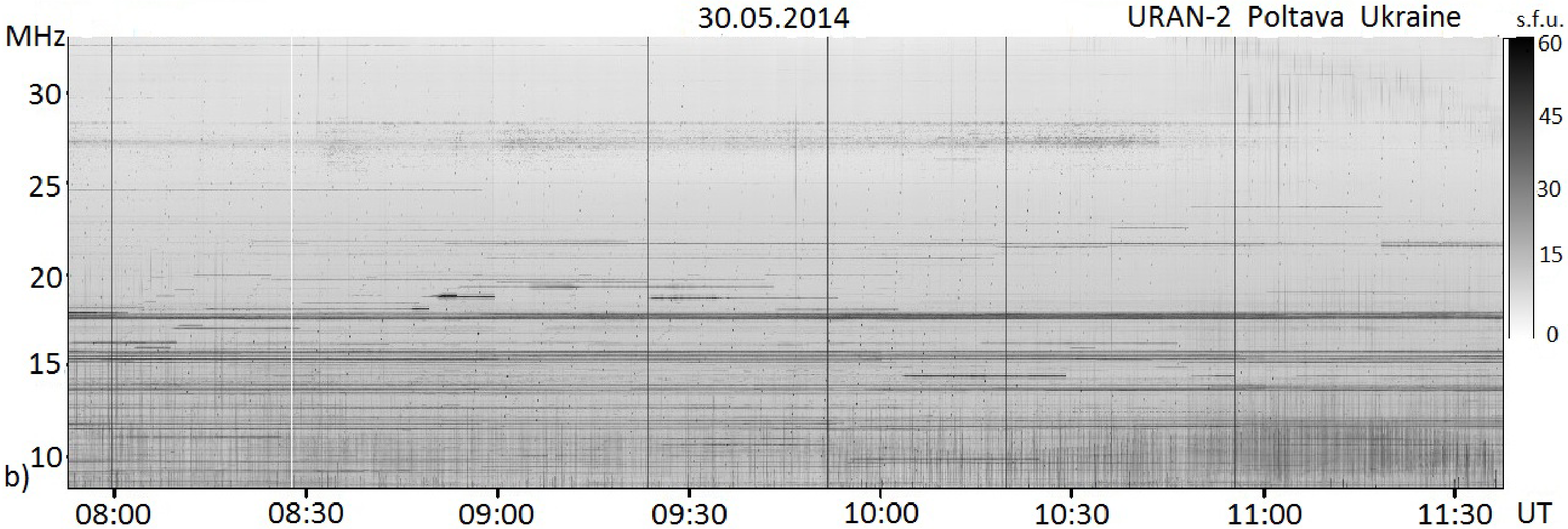}
                            }

\caption{The dynamic spectra of radio emission of the Sun on 28 May (a) and 30 May (b) 2014 in the frequency band of 8\,--\,33 MHz according to the URAN-2 radio telescope.
        }
   \label{F3}
   \end{figure}

\begin{figure}    
   \centerline{\hspace*{0.015\textwidth}
               \includegraphics[width=0.515\textwidth,clip=]{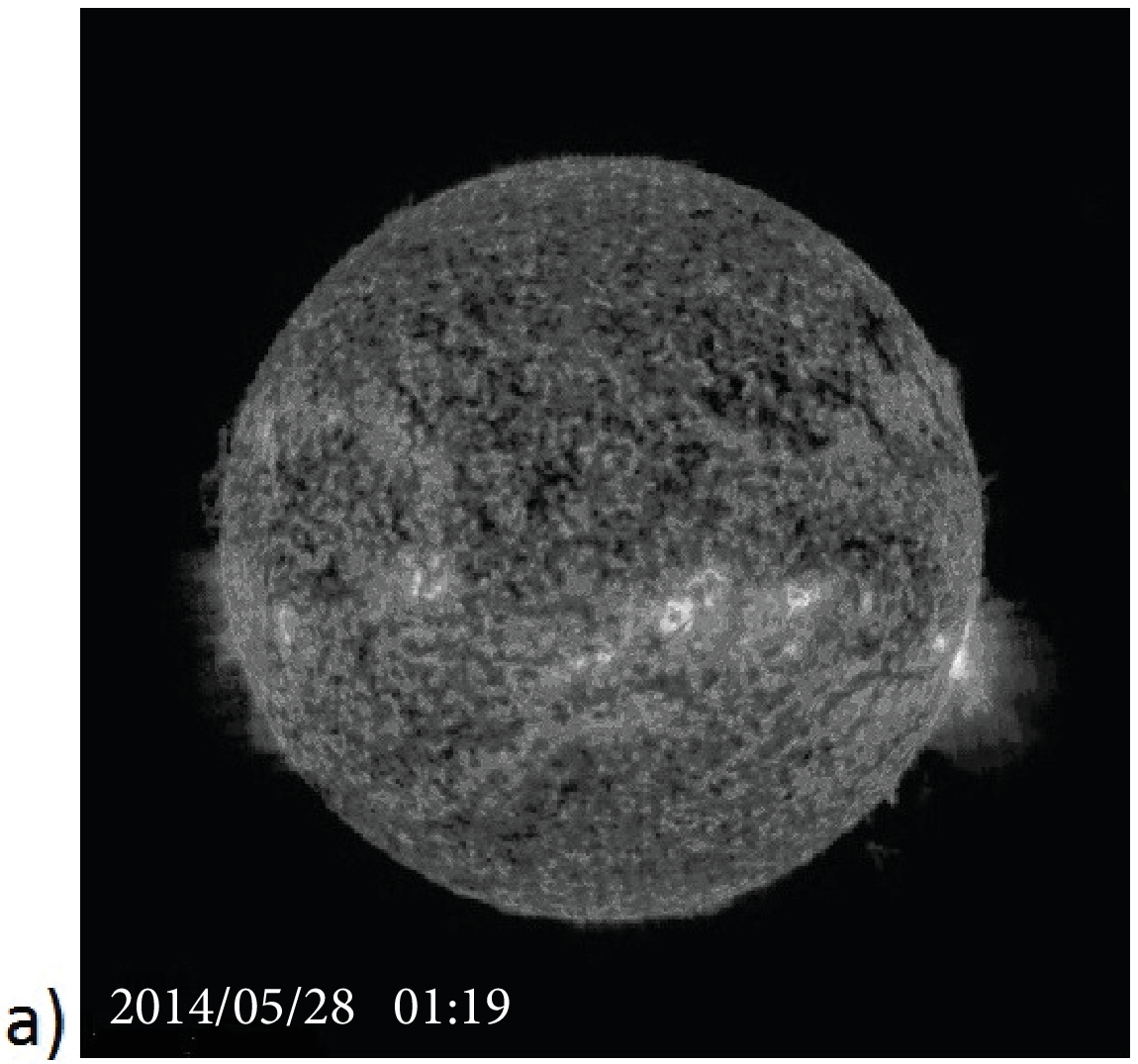}
               \hspace*{-0.01\textwidth}
               \includegraphics[width=0.515\textwidth,clip=]{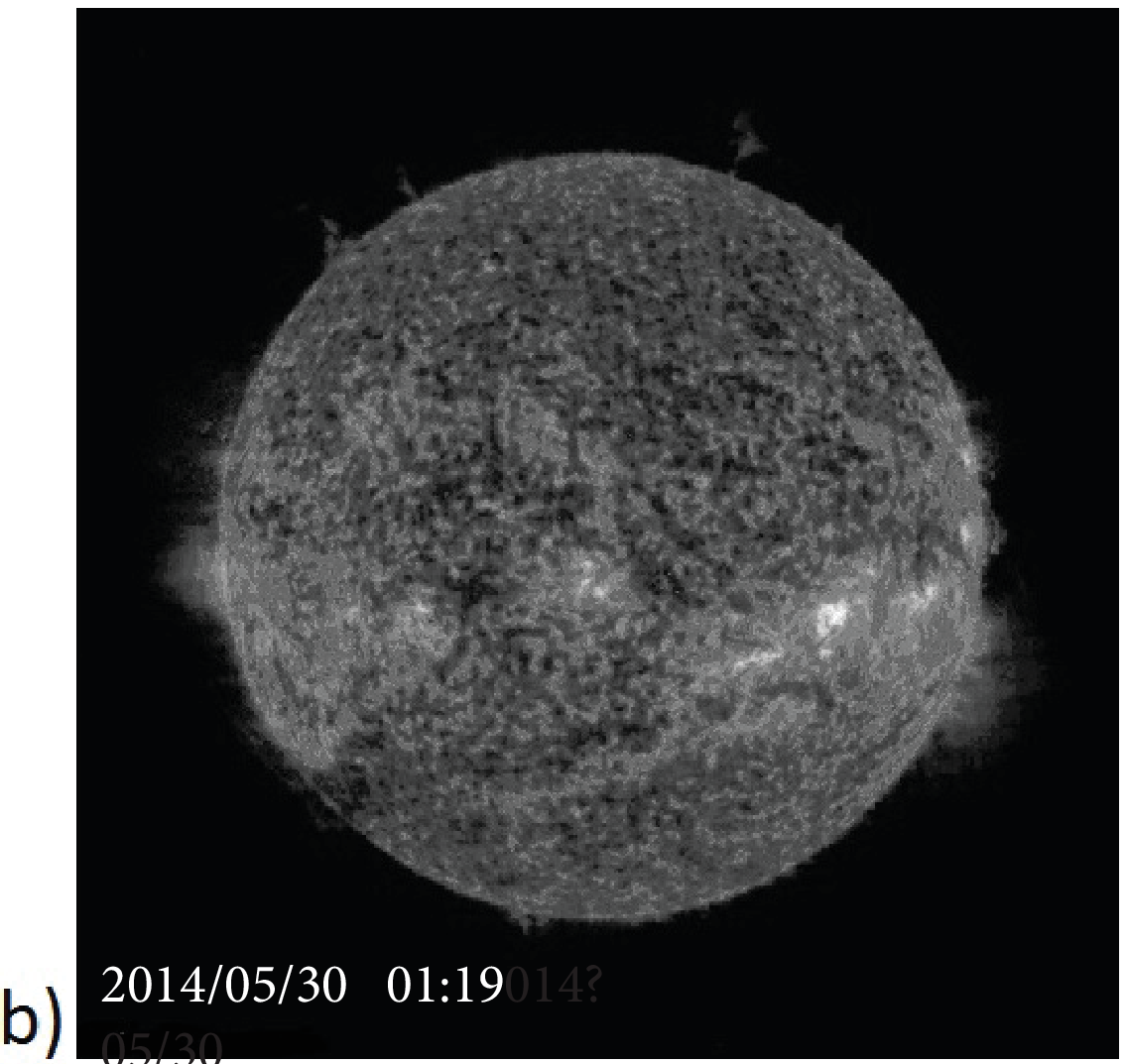}
              }
 \caption{Disk of the Sun on 27 May  and 30 May 2014 according to SOHO (the \emph{Extreme ultraviolet Imaging Telescope} at 304 \AA).
        }
   \label{F4}
   \end{figure}

\begin{figure}    
   \centerline{\hspace*{0.015\textwidth}
               \includegraphics[width=0.515\textwidth,clip=]{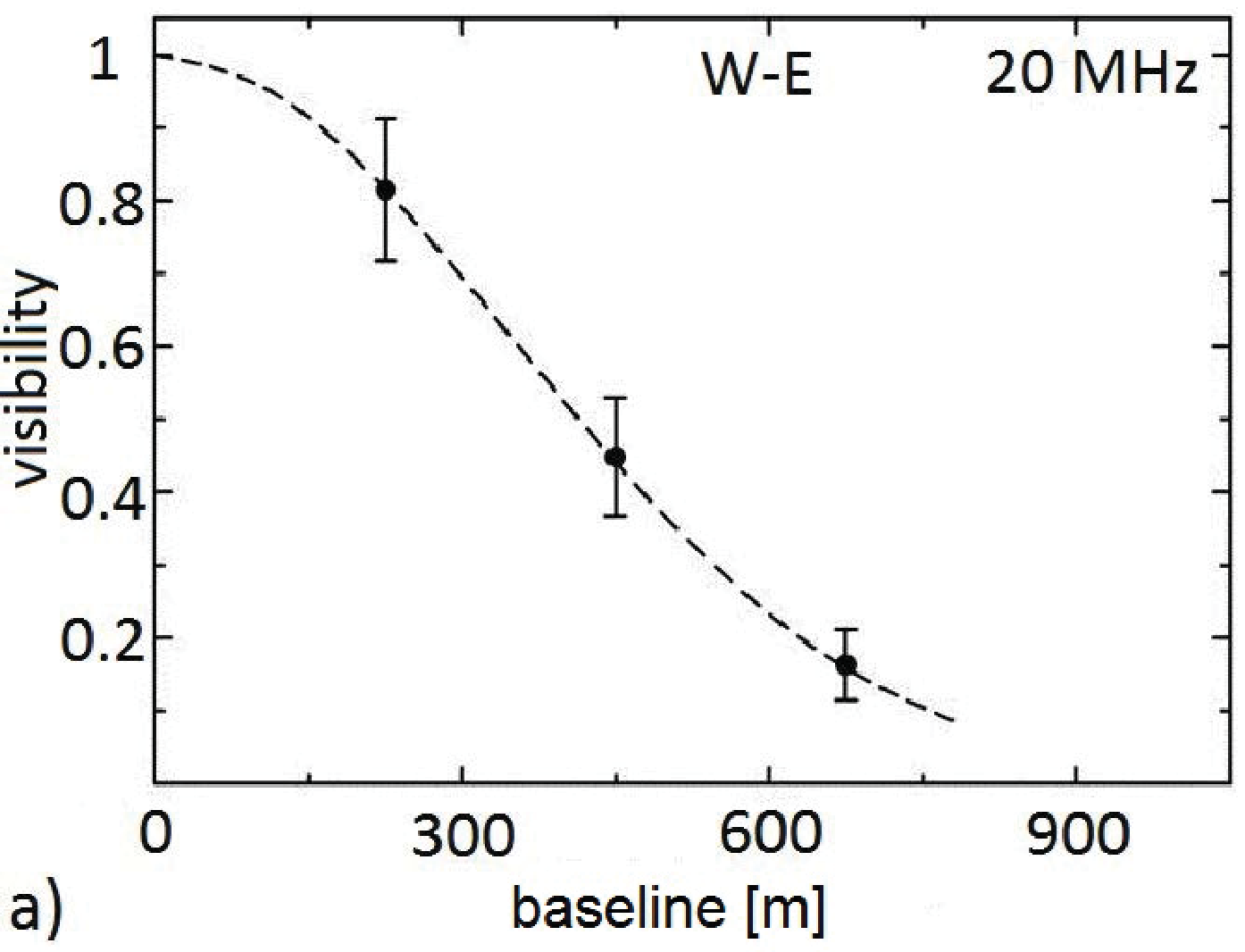}
               \hspace*{-0.01\textwidth}
               \includegraphics[width=0.525\textwidth,clip=]{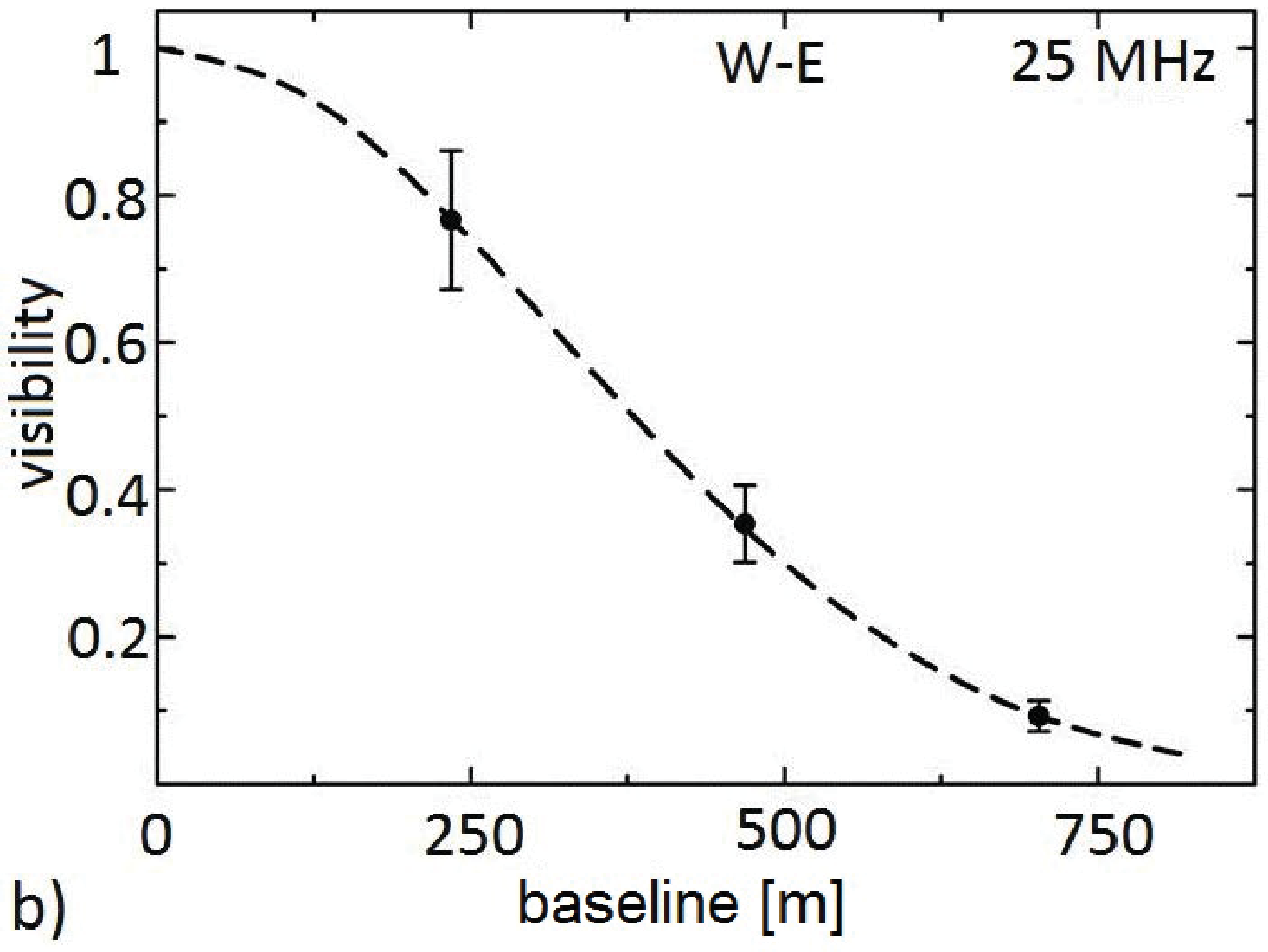}
              }
     \vspace{-0.35\textwidth}   
     \centerline{\Large \bf     
      \hspace{0.0 \textwidth}  \color{white}{(a)}
      \hspace{0.415\textwidth}  \color{white}{(b)}
         \hfill}
     \vspace{0.31\textwidth}    
   \centerline{\hspace*{0.015\textwidth}
               \includegraphics[width=0.505\textwidth,clip=]{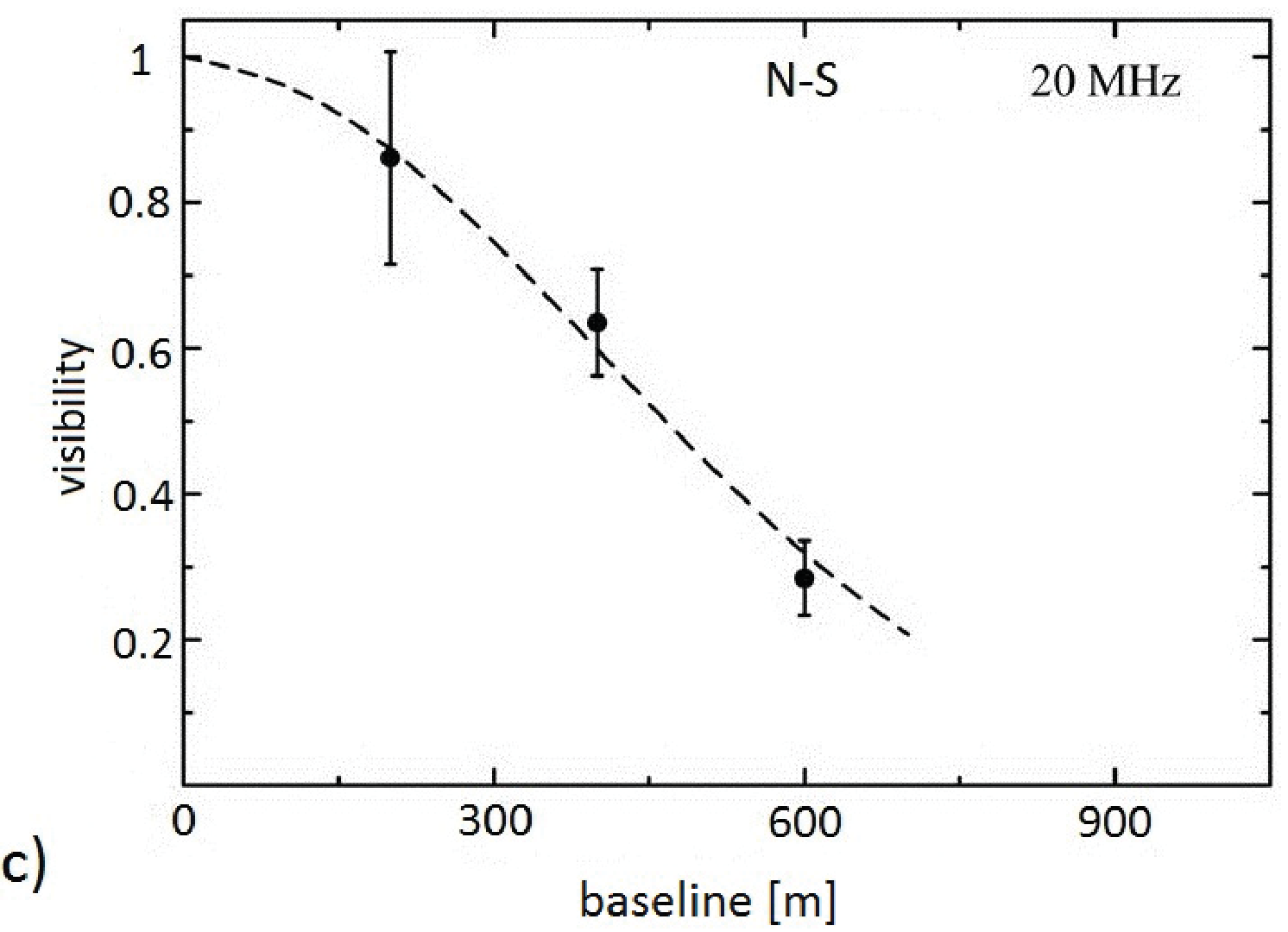}
               \hspace*{-0.01\textwidth}
               \includegraphics[width=0.525\textwidth,clip=]{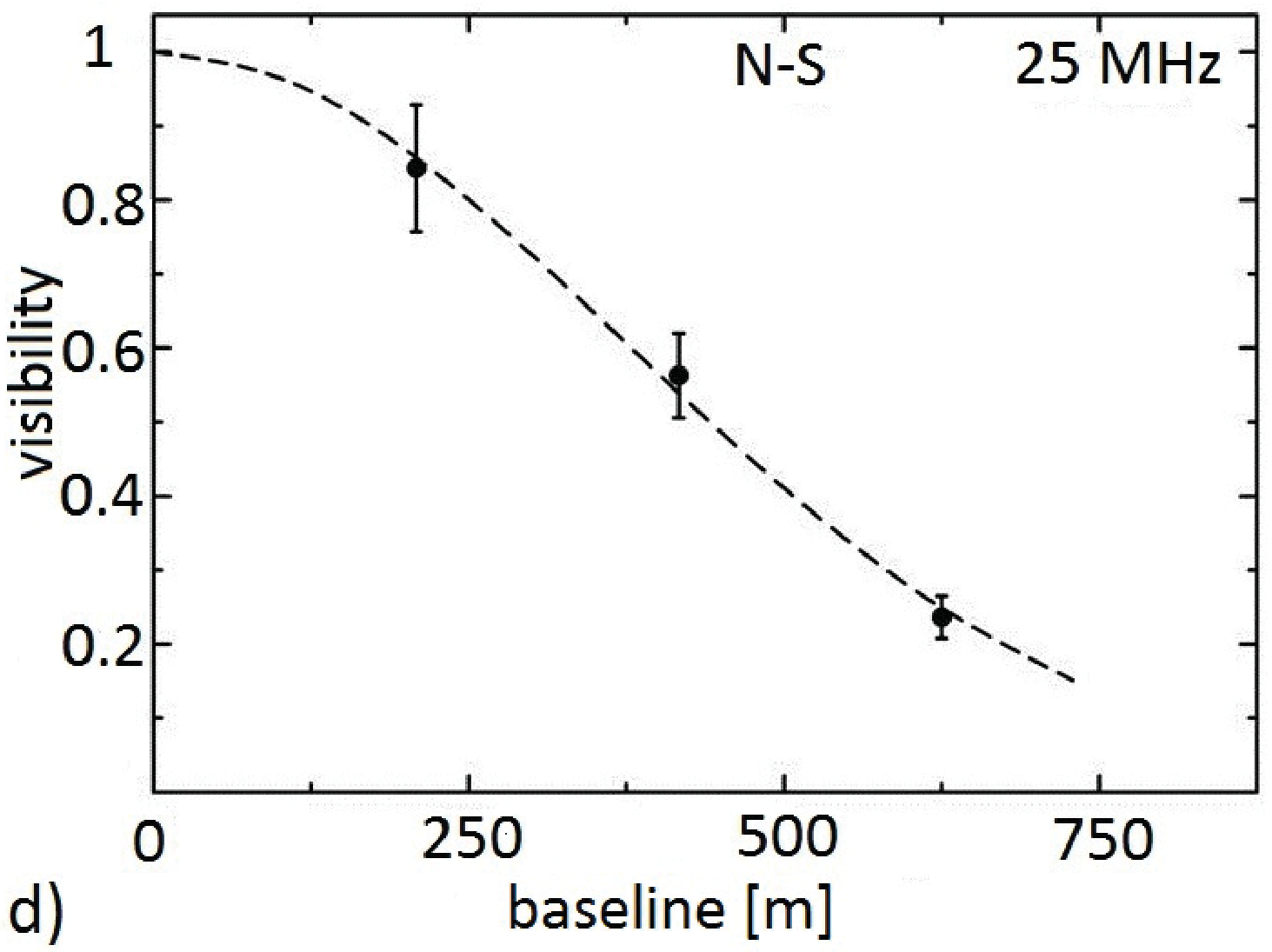}
              }
     \vspace{-0.35\textwidth}   
     \centerline{\Large \bf     
      \hspace{0.0 \textwidth} \color{white}{(c)}
      \hspace{0.415\textwidth}  \color{white}{(d)}
         \hfill}
     \vspace{0.31\textwidth}    

\caption{ Visibility functions of the quiet Sun in East--West (27, 28 and 30 May) (a,b) and North--South (31 May, 2 June) (c,d) directions at 20 and 25 MHz.
        }
   \label{F5}
   \end{figure}

Solar radio emission in the period 27\,--\,30 May 2014 was observed by the interferometer consisting of East--West sections with baselines 225, 450, and 675 m. In the period from 31 May to 2 June the interferometer consisting of North--South sections were used with the same baselines. Thus the sizes of the quiet Sun in the equatorial and polar directions were measured on 27\,--\,30 May and from 31 May to 2 June, correspondingly. Average visibilities were derived from the scans without interferences and close to noon. These values did not change significantly from day to day, so the average visibilities for 27, 28, 30 May (East--West direction) and 31 May, 2 June (North--South direction) are presented in Figure 5.

Supposing Gaussian distributions of the quiet Sun brightness, the curves were derived from the observational data according to Equation 1. Corresponding angular sizes of the quiet Sun in East--West and North--South directions at frequencies 20 and 25 MHz are summarized in Table 1. These sizes are close to those obtained by \citeauthor{Kundu77}  (\citeyear{Kundu77}) in interferometric observations by the radio telescope at Clark Lake, although those observations were carried out in the years of minimum solar activity. At the same time heliograph measurements \citep{Abranin86,Stanislavsky13} gave slightly larger sizes in both directions. Determination of the sizes by the method in which the radio source is passing through the antenna beam as was  reported by \citeauthor{Aubier71}  (\citeyear{Aubier71}), \citeauthor{Sastry94}  (\citeyear{Sastry94}), \citeauthor{Subramanian04}  (\citeyear{Subramanian04}), and \citeauthor{Ramesh06}  (\citeyear{Ramesh06}), gave probably larger sizes. A distinction of sizes can be connected with their dependences on the activity of the Sun. Thus regular observations with different methods are required to define the solar sizes as a function of time and solar activity. Under the assumption that the sizes of the quiet Sun did not change significantly during the week of our observations its compression factor (polar to equatorial size ratio) was 0.89 and 0.84 at 20 and 25 MHz, respectively. These values are close to those obtained by \citeauthor{Hulst47}  (\citeyear{Hulst47}).
Measured fluxes of the quiet Sun were calibrated by the source 3C144 whose flux is well known and equals 3000 Jy at a frequency of 25 MHz \citep{Bobeiko79}. Observations were carried out in such a way that during 20 minutes (two scans) the interferometer beam was pointed at the Sun and after that it followed the source 3C144. The average flux per day was defined on the scan near the local noon. Figure 6 shows that the fluxes before and after the CME were practically the same.

\begin{table}
\caption{ Sizes and fluxes of the quiet Sun at 20 and 25 MHz according to observations on 27 May--2 June 2014.
}
\label{T}
\begin{tabular}{ccccc}     
  \hline                   
    & \multicolumn{2}{c}{E--W direction}& \multicolumn{2}{c}{N--S direction} \\
frequency [MHz] & size [min] & flux [Jy] & size [min] & flux [Jy] \\

  \hline
20 & $55\pm 4$ & $1050\pm 150$ & $49\pm 3$ & $1100\pm 230$ \\
25 & $50\pm 3$ & $1480\pm 130$ & $42\pm 2$ & $1570\pm 190$ \\
  \hline
\end{tabular}
\end{table}

\begin{figure}    
   \centerline{\includegraphics[width=0.7\textwidth,clip=]{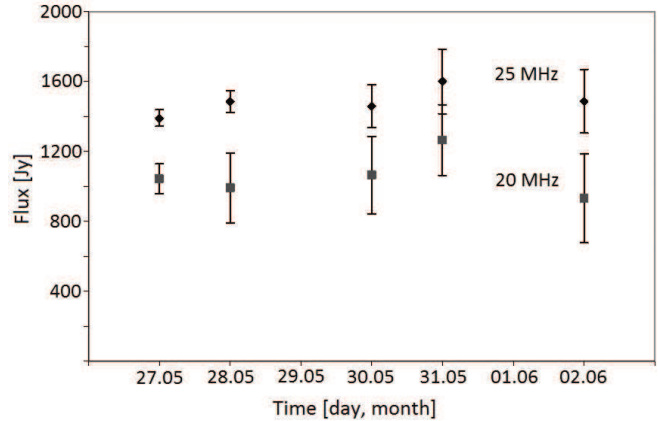}
              }
              \caption{ Fluxes of the quiet Sun at frequencies 20 and 25 MHz from 27 May to 2 June 2014.
                      }
   \label{F6}
   \end{figure}

The measured fluxes were comparable to the fluxes obtained by \citeauthor{Erickson77}  (\citeyear{Erickson77}) at the radio telescope at Clark Lake and significantly smaller than those from observations by Arecibo \citep{Aubier71} and significantly larger than fluxes registered by the UTR-2 and URAN-2 radio telescopes \citep{Abranin86,Brazhenko12,Stanislavsky13}. In the last cases observations were conducted at minimum of solar activity so the difference seems understandable.

\section{Discussion}
\label{S-dis}

The brightness temperature of the quiet Sun is defined by the equation \citep{Aubier71}

\begin{equation}  {{T}_{\rm b}}=5.5\times {{10}^{29}}\frac{{{\lambda }^{2}}S}{{{\theta }_{\rm W}}{{\theta }_{\rm N}}}
   \end{equation}
where S is the flux [ Wm$^{-2}$Hz$^{-1}$], ${{\theta }_{\rm W}}$,
${{\theta }_{\rm N}}$ are diameters [minutes] in equatorial and polar directions and $\lambda$  is the wavelength [m]. Using the average fluxes and sizes obtained at frequencies 20 and 25 MHz, we get for brightness temperatures of the quiet Sun ${{T}_{\rm b}}=5.1\times {{10}^{5}}$ K and ${{T}_{\rm b}}=5.7\times {{10}^{5}}$ K at these frequencies. These values, as well as temperatures derived by other authors, are noticeably smaller than the temperature of the coronal plasma, which usually is supposed to be equal to ${{T}_{\rm e}}={{10}^{6}}$ K.  Such a difference can be explained by scattering of electromagnetic waves on the density inhomogeneity of coronal plasma. Such a supposition was stated first by \citeauthor{Aubier71}  (\citeyear{Aubier71}) and then was discussed many times \citep{Abranin86,Thejappa92,Subramanian04,Thejappa08}. \textbf{\citeauthor{Thejappa08}  (\citeyear{Thejappa08}) recently studied in details an influence of processes of scattering in the solar corona on brightness temperatures and sizes of the quiet Sun at low frequencies by the Monte Karlo method supposing Kolmogorov and flat spectra of random density fluctuations. They obtained that brightness temperatures of the quiet Sun at 30 MHz can be as low as $1.5\times {10}^{5}$\,--\,$4.5\times {10}^{5}$ K and it can be even much less at lower frequencies. Thus our brightness temperatures 5\,--\,6$\times {10}^{5}$ K can be explained by scattering in the solar corona as stated by other authors \citep{Aubier71,Abranin86,Thejappa92}. It is  worth to say that there exists another point of view \citep{Sastry94,Subramanian04,Ramesh06}, according to which a high level of density fluctuations leads to very high sizes of the quiet Sun. But \citeauthor{Thejappa08}  (\citeyear{Thejappa08}) using Monte Karlo method found that obtained sizes are close to observed ones. We consider that choosing of level of density fluctuations, their inner and outer scales, and their degree of asymmetry can give reasonable values of brightness temperatures and sizes at different frequencies and vice versa. We plan to observe the quiet Sun in the whole working frequency band of UTR-2 radio telescope, 8-32 MHz, and using different methods including Monte Karlo method to find parameters of density fluctuations in the solar corona at heights 0.5\,--\,2 R$_\odot$.} Concerning the sizes obtained, it is worth  pointing out that the radii of the quiet Sun are close to the distances from the Sun at which the local plasma frequency is equal to the corresponding observed frequency of radio emission in the Baumbach--Allen model \citep{Allen47} (Figure 7). At the same time we see (Figure 7) that observational points are situated notably below the Newkirk model $n(r) = {n_0}{10^{4.32{R_\odot}/r}}$ (${{n}_{0}}=4.2\times {{10}^{4}}c{{m}^{-3}}$, and $r$ is the distance from the Sun) \citep{Newkirk61}, which agreed with the Mann's model \citep{Mann99} for coronal plasma temperature,  ${{T}_{\rm e}}=1.9\times {{10}^{6}}$ K.

\begin{figure}    
   \centerline{\includegraphics[width=0.7\textwidth,clip=]{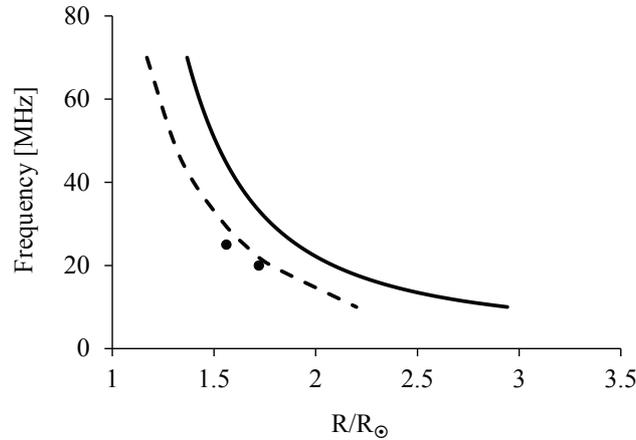}
              }
              \caption{ Baumbach--Allen (dashed) and Newkirk (solid) models and radii of the quiet Sun according to observations presented.
	                      }
   \label{F7}
   \end{figure}

\section{Conclusion}
\label{S-Conclusion}

Interferometric observations of the quiet Sun with the UTR-2 radio telescope showed that they can be used for the determination of the quiet Sun parameters such as fluxes, sizes, and brightness temperatures at heights 0.5--2R$_\odot$. From this point of view it is of current interest to regularly observe the quiet Sun in the decameter range. During the first observations in interferometric session at the radio telescope UTR-2 it was shown that:

- Gaussian distribution of the quiet Sun intensity is a good approximation;

- brightness temperatures of the quiet Sun are ${{T}_{\rm b}}=5.1\times {{10}^{5}}$ K and ${{T}_{\rm b}}=5.7\times {{10}^{5}}$ K at 20 and 25 MHz, correspondingly, and they are smaller than the temperature of the coronal plasma;

- obtained radii of the quiet Sun at 20 and 25 MHz are close to distances from the Sun of corresponding local plasma frequencies in the Baumbach--Allen solar corona;

- compression factor of the quiet Sun is about 0.85\,--\,0.9 in the decameter range;

- radio fluxes of the quiet Sun at 20 and 25 MHz are practically the same on the days before and after a CME.

\begin{acks}
 The work was partially financed in the frame of FP7 project SOLSPANET (FP7-PEOPLE-2010-IRSES-269299).
\end{acks}

\section* {Disclosure of Potential Conflicts of Interest} The authors declare that they have no conflicts of interest.

\bibliographystyle{spr-mp-sola}
\bibliography{sun3}

\end{article}

\end{document}